\documentclass[prl,twocolumn,nofootinbib]{revtex4}
\usepackage{graphicx, epsfig}
\usepackage{color}
\usepackage{mathrsfs}
\usepackage{bm}
\usepackage{amsmath,amssymb}
\usepackage{mathrsfs}
\usepackage[caption=false]{subfig}
\usepackage{hyperref}
\usepackage[normalem]{ulem}
\usepackage{mathtools}
\usepackage[x11names]{xcolor}

\definecolor{fashionfuchsia}{rgb}{0.96, 0.0, 0.63}
\colorlet{no_so_fashion_purple}{blue!50!red}

\newcommand{\be}{\begin{equation}}
\newcommand{\ee}{\end{equation}}
\newcommand{\ba}{\begin{eqnarray}}
\newcommand{\ea}{\end{eqnarray}}

\newcommand{\nn}{\nonumber}

\newcommand{\calE}{{\cal E}}

\begin{document}
\title{Electric strings in non-Abelian theories}
\author{Tanmay Vachaspati}
\affiliation{
$^*$Physics Department, Arizona State University, Tempe,  Arizona 85287, USA.
}

\begin{abstract}
We find electric string solutions in Yang-Mills-Higgs theory.
\end{abstract}

\maketitle

Cosmic strings have attracted a lot of attention for nearly half a century since their discovery~\cite{Nielsen:1973cs,Vilenkin:2000jqa}. Such strings carry magnetic flux 
and are analogous to vortices in superconductors. In contrast, strings that carry 
electric flux are expected to lead to confinement in QCD and classical solutions
corresponding to electric strings are not known. 
Even if electric strings were to exist as classical solutions in a non-Abelian gauge 
theory, because gauge excitations (``gluons'') are massless and carry non-Abelian 
charge one might expect that rapid production of gluons by the 
Schwinger process~\cite{Schwinger:1951nm,Cardona:2021ovn}
would dissipate such strings . In view of these expectations it is surprising that electric 
string solutions do exist in certain non-Abelian gauge theories and are protected against 
the Schwinger process.

An essential element in constructing an electric string solution\footnote{
Some early work on classical solutions in non-Abelian gauge theories can
be found in Ref.~\cite{Jackiw:1978zi,Huang:1980bz}.
}
is that a non-Abelian 
electric field does not uniquely specify a gauge equivalent class of gauge fields. As
Brown and Weisberger (BW) showed~\cite{Brown:1979bv}, there is a one parameter 
family of gauge inequivalent gauge fields that all result in the same electric field. 
Unlike a uniform non-Abelian electric field produced in analogy with Maxwell theory, 
BW gauge fields have been shown to be immune from decay due to Schwinger pair 
production~\cite{Vachaspati:2022ktr}.
This suggests the question: can there be string solutions containing BW gauge
fields? 
In pure non-Abelian gauge theory, BW gauge fields do not solve the classical
equations of motion. Instead they require external current and charge densities.
Such external sources may arise due to quantum effects -- after all the 
classical equations are expected to get modified due to the backreaction
of quantum fluctuations -- or they may be due to other fields in the system.
Here we consider $SU(2)$ gauge theory with a scalar field in the fundamental
representation. The same solution can be embedded in models with larger
gauge group that have an $SU(2)$ 
subgroup~\cite{Vachaspati:1992pi,Barriola:1993fy,Preskill:1992bf}. The Lagrangian
under consideration is,
\be
L = - \frac{1}{4}W_{\mu\nu}^a W^{\mu\nu a} + | D_\mu\Phi |^2 - V(\Phi)
\label{Lag}
\ee
where
\be
W_{\mu\nu}^a = \partial_\mu W_\nu^a - \partial_\nu W_\mu^a + g \epsilon^{abc}W_\mu^b W_\nu^c
\ee
\be
D_\mu \Phi = \partial_\mu \Phi - i \frac{g}{2} W_\mu^a \sigma^a \Phi
\ee
\be
V(\Phi ) = m^2 |\Phi |^2 + \lambda |\Phi |^4
\ee
where $\sigma^a$ are the Pauli spin matrices.
The Lagrangian actually has an $SU(2)\times U(1)$ symmetry but only the 
$SU(2)$ is gauged, while the $U(1)$ is global. (This corresponds to the
$\sin^2\theta_w =0$ limit of the electroweak model.)

The equations of motion are
\be
(\partial_\nu + g \epsilon^{abc} W_\nu^b)W^{\mu\nu c} = 
i \frac{g}{2} \left ( \Phi^\dag \sigma^a D^\mu\Phi - h.c. \right )
\label{Weom}
\ee
\be
D^\mu D_\mu \Phi + m^2 \Phi + 2 \lambda | \Phi |^2 \Phi =0 .
\label{Phieom}
\ee

\begin{figure}
\includegraphics[width=0.5\textwidth,angle=0]{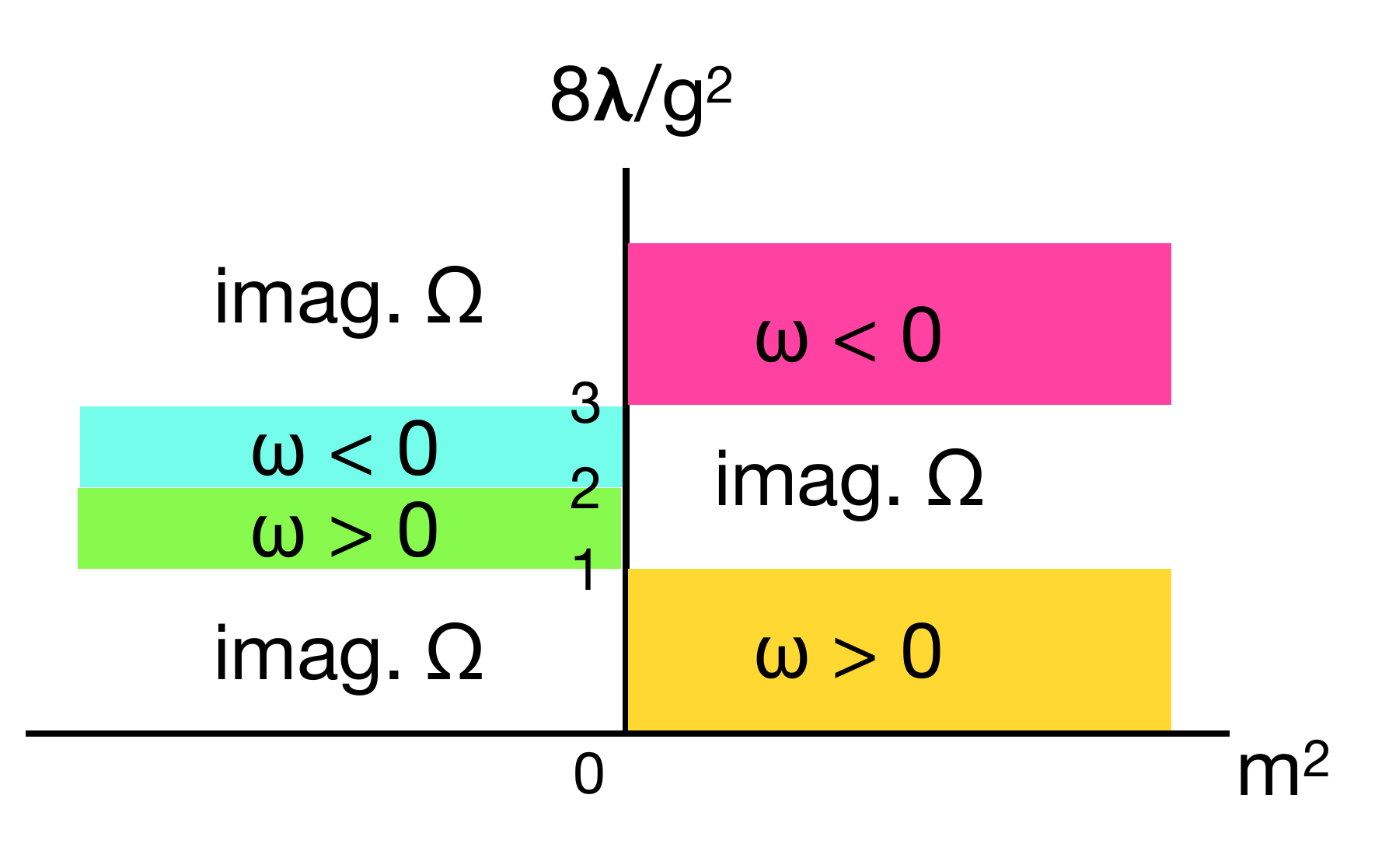}
 \caption{Constraints on the parameters in the $m^2$-$8\lambda/g^2$ plane. The unshaded
 regions give imaginary $\Omega$ and are not allowed. 
 The solution is only valid in the regions of parameter space where $\omega > 0$.
 }
\label{constraints}
\end{figure}

To find an electric field solution to the equations of motion, we first write the
BW gauge field in temporal gauge~\cite{Vachaspati:2022ktr},
\be
W_0^a = 0, \ \ W^\pm_\mu = -\frac{\epsilon}{g}  e^{\pm i \Omega t} f(r) \, \partial_\mu z , \ \ 
W^3_\mu =0, 
\label{stringW}
\ee
where $W^\pm_\mu = W^1_\mu \pm i W^2_\mu$, and we have introduced a cylindrical profile 
function, $f(r)$, with $f(0)=1$.
This gauge field needs sources as given by the right-hand side of \eqref{Weom}. Requiring that
$\Phi$ provides the currents that will produce the gauge field in \eqref{stringW}
allows us to construct $\Phi$, which we have to make sure also 
satisfies \eqref{Phieom}. Some algebra shows that the required form of $\Phi$ is,
\be
\Phi = \frac{\epsilon}{g} \sqrt{\frac{\Omega}{\omega}} \, f(r)
\begin{pmatrix} z_1 e^{+i\omega t} \\ z_2 e^{-i\omega t} \end{pmatrix}
\label{stringPhi}
\ee
where $W^\pm_\mu = W^1_\mu \pm i W^2_\mu$ and $|z_1|^2+|z_2|^2=1$. 
Without loss of generality we take $\Omega > 0$ while $\omega$ can be positive or negative. 
The $\Phi$
field can then provide suitable sources for the gauge field as well as satisfy its classical
equation of motion provided we take
\ba
\Omega &=& m \left [ 4(1-4\lambda/g^2)^2-1 \right ]^{-1/2} \label{Omsol} \\
\omega &=&  2(1-4\lambda/g^2) \Omega \label{omsol}
\ea
with restrictions that are necessary for $\Omega$ to be real, namely, for $m^2 > 0$, 
we need $0 < \lambda < g^2/8$ ($\omega > 0$), or $\lambda > 3g^2/8$ ($\omega < 0$),
and for $m^2 < 0$, $g^2/8 < \lambda < g^2/4$ ($\omega > 0$) and
$g^2/4 < \lambda < 3 g^2/8$ ($\omega < 0$). 
An explicit check (see Supplemental Material) shows that the solution is only valid for 
$\omega > 0$, {\it i.e.} for $\lambda < g^2/4$. In what follows we will assume $\omega > 0$.
These constraints are summarized in Fig.~\ref{constraints}.

The profile function satisfies the equation,
\be
 f'' + \frac{f'}{r} + \Omega^2 \left ( 1 - \frac{\epsilon^2 }{2\omega  \Omega} f^2 \right ) f = 0.
 \label{hfdiffeq2}
\ee
with boundary conditions $f(0)=1$, $f'(0)=0$. 

\begin{figure}
\includegraphics[width=0.4\textwidth,angle=0]{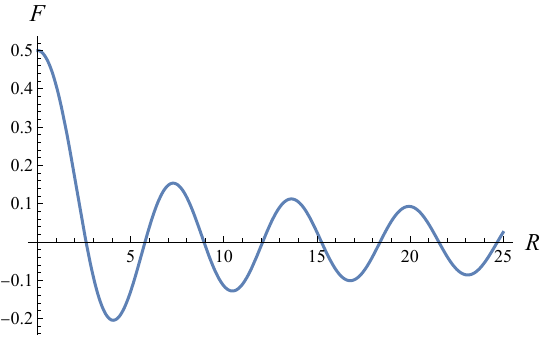}
 \caption{$F(R)$ vs. $R$ for $F(0)=0.5$.}
\label{plotprofile}
\end{figure}

For $\epsilon^2 \ll 2\omega \Omega$ the 
solution is closely approximated by the zeroth order Bessel function, $J_0(\Omega r)$.
For $\epsilon^2 /2\omega\Omega > 1$, there is no well-behaved solution. 
We define a rescaled profile $F(r)\equiv \epsilon f(r)/\sqrt{2\omega  \Omega}$ and
a rescaled coordinate $R=\Omega r$.
A numerical solution for $F(R)$ vs. $R$ is shown in Fig.~\ref{plotprofile}. 
The asymptotic form of the solution is approximately described by the 
asymptotic form of the Bessel function,
\be
F(R) \sim F(0) \sqrt{\frac{2}{\pi R}} \, \cos \left ( R-\frac{\pi}{4} \right ), \ \ R \to \infty .
\label{asympF}
\ee

\begin{figure}
\includegraphics[width=0.4\textwidth,angle=0]{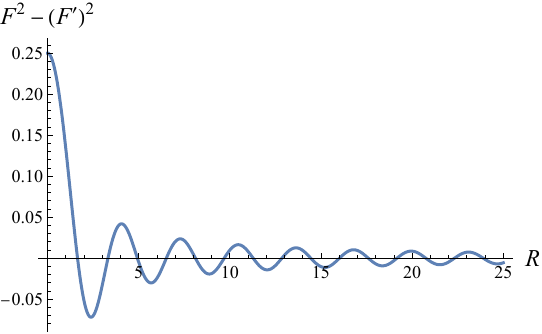}
 \caption{$F(R)^2-F'(R)^2$ vs. $R$ for $F_0=0.5$. The field strength is
electric-like where $F(R)^2-F'(R)^2$ is positive and magnetic-like where
$F(R)^2-F'(R)^2$ is negative.}
\label{plotLag}
\end{figure}

The field strength for the solution is
\ba
W_{\mu\nu}^\pm &=& -\frac{\epsilon}{g} e^{\pm i\Omega t} \bigl [ i \Omega  f(r) 
(\partial_\mu t \, \partial_\nu z - \partial_\nu t \, \partial_\mu z) \nn \\
&& \hskip 1.75 cm
+ f'(r) (\partial_\mu r \, \partial_\nu z - \partial_\nu r \, \partial_\mu z) \bigr ]
\label{Wmunu}
\ea
and $W_{\mu\nu}^3=0$. An $SU(2)$ gauge rotation by
\be
U = e^{i\sigma^1 \Omega t/2} e^{- i\sigma^1 \pi/4} e^{ -i \sigma^2 \pi/4}
\ee
brings the field strength to the form,
\be
W_{\mu\nu}' = U W_{\mu\nu} U^\dag
\ee
where $W_{\mu\nu}= W_{\mu\nu}^a \sigma^a$. This gives static field strengths of the
form in~\cite{Brown:1979bv} (if we set $f=1$),
\ba
&&
W_{\mu\nu}^{1\, \prime} =0 
\label{WBW1} \\
&&
W_{\mu\nu}^{2\, \prime}  = -\frac{\epsilon}{g} f'(r) ( \partial_\mu r \, \partial_\nu z - \partial_\nu r \, \partial_\mu z)
\label{WBW2} \\
&&
W_{\mu\nu}^{3\, \prime}  = -\frac{\epsilon}{g}\Omega f(r) ( \partial_\mu t \, \partial_\nu z - \partial_\nu t \, \partial_\mu z)
\label{WBW3}
\ea
Then the electric field is in the third $SU(2)$ direction and the spatial $z-$direction, while the magnetic
field is in the second $SU(2)$ direction and in the spatial azimuthal direction. The structure of the solution
consists of a tube of electric field along $z$, wrapped by magnetic field along the azimuthal direction
$\hat \phi$, 
which is then within a sheath of electric field in the $-z$ direction, wrapped in magnetic field 
in the $-\hat\phi$ direction, and so 
on. The distinction between electric and
magnetic fields is frame-dependent and so we calculate the Lorentz invariant 
$-W_{\mu\nu}^aW^{\mu\nu a} \propto F^2 - F'^2$ in Fig.~\ref{plotLag}, confirming the
alternating sequence of electric and magnetic fields.

The gauge transformation $U$ when applied to $\Phi$ gives,
\be
\Phi' = U \Phi = \frac{\epsilon}{g} \sqrt{\frac{\Omega}{2\omega}} \, f(r)
\begin{pmatrix}
z_1' e^{i\omega' t}-z_2' e^{-i\omega' t} \\
z_1' e^{i\omega' t} + z_2' e^{-i\omega' t}
\end{pmatrix}
\label{Phi'}
\ee
where $\omega'= \omega+\Omega/2$ and $z_1'=z_1 e^{-i\pi/4}$, $z_2'=z_2 e^{i\pi/4}$.

\begin{figure}
\includegraphics[width=0.4\textwidth,angle=0]{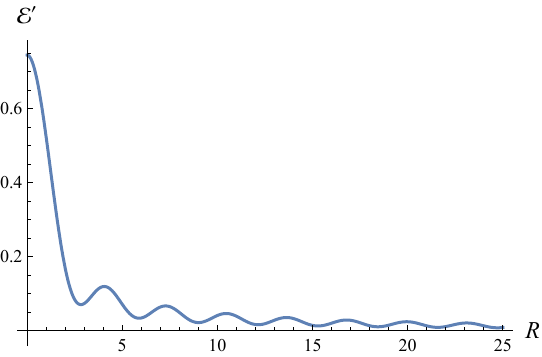}
 \caption{$\calE'$ vs. $R$ for $F_0=0.5$ and $\kappa =1.5$.}
\label{calEplot}
\end{figure}

The energy density, ${\cal E}$ of the solution can be calculated from the expression,
\be
\calE = \frac{1}{2} (W_{0i}^a )^2 +  \frac{1}{4} (W_{ij}^a )^2 + 
|D_t\Phi |^2 + |D_i\Phi |^2 + V(\Phi )
\ee
In terms of rescaled variables, and restricting to $\omega > 0$,
\ba
\calE' \equiv \frac{g^2}{2\omega  \Omega^3}\calE
&=& \left ( \frac{1}{2} + \frac{1}{\kappa} \right ) F'^2 +
\left ( \frac{1}{2} + 2\kappa - \frac{1}{\kappa} \right ) F^2 \nn \\
&&
+  \frac{1}{2} \left ( \frac{1}{2} + \frac{1}{\kappa} \right ) F^4
\label{energy'}
\ea
where (see \eqref{omsol}) $0 < \kappa \equiv \omega /\Omega < 2$.
In Fig.~\ref{calEplot} we show an example of $\calE'$ vs. $R$.

The slow fall off $\propto 1/\sqrt{R}$ of the gauge fields in \eqref{asympF} implies that
$\calE \propto 1/R$ and that the energy per unit length, $\mu$, diverges linearly with
radial distance. Hence the string is not localized as in a Nielsen-Olesen string but
is more like a global string that has a logarithmically divergent energy per
unit length, or like a global monopole with linearly divergent energy~\cite{Vilenkin:2000jqa}.
We evaluate $\mu$ numerically by integrating over $R \in [0,R_c]$ 
where $R_c$ is a radial cutoff to get,
\ba
\mu(R_c) &=& 2\pi \int_0^{R_c} dR \, R \, \calE'  \nn \\
&\approx& \left [ \left ( \frac{1}{2} + \frac{1}{\kappa} \right ) 0.54 +
\left ( \frac{1}{2} + 2\kappa - \frac{1}{\kappa} \right ) 0.54 \right ] R_c \nn \\
&&
+  \frac{1}{2} \left ( \frac{1}{2} + \frac{1}{\kappa} \right ) 0.09\, \ln (R_c), \ \ (\omega > 0)
\ea
In the $m^2 < 0$ case, we should add a constant piece to the potential so that
$V=0$ at its minimum.
In that case, however, the solution still has $|\Phi | \to 0$ in the asymptotic region. 
Since the true vacuum has $|\Phi | \ne 0$, the solution has divergent energy per 
unit length and the divergence will go as $R_c^2$ instead of $R_c$.

Our electric string solution is with a scalar field in the fundamental
representation. This leads to the question whether scalar fields in other
representations can also provide suitable sources for the gauge fields
in \eqref{stringW}. We have examined the
case of a scalar field in the adjoint representation, $\phi^a$, and can show that a solution
does not exist. Briefly, the gauge field equation is $D_\nu W^{\mu\nu a} = j^{\mu a}$
where the current $j^{\mu a}$ is now due to the adjoint scalar and has a form such
that $\phi^a j_\mu^a =0$ for every value of the index $\mu$. This constraint then
requires $\phi^a D_\nu W^{\mu\nu a} =0$ for every $\mu$, where the gauge
field is given in \eqref{stringW}. The constraint is strong enough that it fixes
the form of $\phi^a$. Then we can calculate $j_\mu^a$ using $\phi^a$ and we see that it 
does not satisfy the gauge field equation of motion: $ j^{\mu a}=D_\nu W^{\mu\nu a}$.

Another interesting question is if electric string solutions exist in the electroweak
model. The difference between the model in \eqref{Lag} and the bosonic sector
of the electroweak model
is that the global $U(1)$ symmetry of \eqref{Lag} is now
gauged with coupling $g'$ and $\Phi$ has non-vanishing $U(1)$ gauge charge 
(known as ``hypercharge''). Then the $U(1)$ hypercharge current is,
\be
j_\mu^Y = i \frac{g'}{2} \left ( \Phi^\dag D_\mu \Phi - h.c. \right )
\ee
Using \eqref{stringW} and \eqref{stringPhi} we find,
\ba
j_\mu^Y &=& -\frac{g' \epsilon^2}{g^2} \Omega f^2 (|z_1|^2-|z_2|^2) \partial_\mu t 
\nn \\
&& 
-\frac{g' \epsilon^3}{2g^2} \frac{\Omega}{\omega} f^3 
\left [ z_1 z_2^* e^{i(\Omega +2\omega)t} + c.c. \right ] \partial_\mu z,
\label{jmuY}
\ea
which shows that in general there is a hypercharge charge density as well as
a three current. If $j_\mu^Y$ is non-vanishing, then necessarily the hypercharge
gauge field, $Y^\mu$, is non-vanishing because it satisfies,
\be
\partial_\nu Y^{\mu\nu} = j^{Y\mu}.
\ee
Our solution does not include a hypercharge component and so the question is
if we can choose parameters such that $j_\mu^Y=0$. The $\mu = 0$ component
can be made to vanish by choosing $|z_1|^2=|z_2|^2$. 
The $\mu=3$ component is time dependent except if $\Omega + 2\omega=0$, which 
corresponds to $\omega = -\Omega/2 < 0$. Since the solution is only valid for $\omega >0$,
it does not hold in the electroweak model with $Y_\mu=0$.
An alternate possibility is that there exist solutions with non-vanishing hypercharge
gauge field $Y_\mu$ and then we do not need to require that $j_\mu^Y$ vanish.
We have not been able to construct such solutions.

An important feature of the gauge field in the solution is that it is stationary.
In other words, consider perturbations of only the gauge field,
\be
W_\mu^\pm = A_\mu^\pm + e^{\pm i \Omega t} Q_\mu^a, \ \ 
W_\mu^3=  Q_\mu^3, \ \ 
\Phi = \Phi_0
\ee
where $A_\mu^a$, $\Phi_0$ denote the electric string solution. Then the action for 
$Q_\mu^a$ does not contain any terms that are explicitly time-dependent and hence 
the solution is protected from decay to
Schwinger pair production of gauge field excitations as shown in Ref.~\cite{Vachaspati:2022ktr}. 
The inclusion of the scalar field, $\Phi$, does not make any difference to the
analysis in Ref.~~\cite{Vachaspati:2022ktr} because the quadratic order interaction between $\Phi$
and $Q_\mu^a$ is simply $g^2 |\Phi |^2 (Q_\mu^a )^2/4$ and $|\Phi |^2$ is
time-independent. (The linear order terms vanish because the solution obeys
the classical equations of motion.) Fluctuations of the scalar field $\Phi$ can
indeed get excited by the time dependence of the solution, and this means
that Schwinger pair production of $\Phi$ excitations will occur. (This is similar
to the Schwinger pair production of quarks on QCD strings.) In the limit
of large mass parameter $m$, the Schwinger pair production of $\Phi$ will
be suppressed.

A Maxwell electric field  for example, with $W_z^3 = -E t$ and all other components 
zero, is a classical solution of the SU(2) pure gauge theory. However, the Schwinger process
for gluons in this background is non-vanishing at all momentum scales~\cite{Cardona:2021ovn}
and the Maxwell electric field will decay and evolve into another configuration. Since
a BW electric field is stable to the Schwinger process, it is likely that it is the final
state. In other words, if initially we start with a Maxwell electric field, it may evolve 
into a BW electric field.

In Ref.~\cite{Pereira:2022lbl} the {\it classical} stability of a homogeneous electric field of the
BW type was analyzed. Several unstable modes were found for the homogeneous
configuration. For the electric string, these unstable modes will be suppressed
due to the $|\Phi |^2 (Q_\mu^a)^2$ term, since this term provides an effective 
mass to the gauge excitations wherever $|\Phi |^2$ is non-zero. The solutions
with $m^2 < 0$ will almost certainly be unstable since $\Phi$ approaches the
unstable vacuum, $|\Phi |=0$, in the asymptotic region.
We plan to carry out a detailed stability analysis in a future publication.

We close with a speculative remark.
The solution we have found may be of interest in the context of pure non-Abelian
gauge theories. In that case, the field $\Phi$ must arise as an effective degree of
freedom due to quantum backreaction in the classical equations of motion.
In this connection, it has been conjectured that magnetic monopoles at strong 
coupling transform as a scalar degree of freedom in the fundamental representation 
of a dual symmetry group~\cite{Goddard:1976qe}.
While there are no classical monopole 
solutions in pure non-Abelian gauge theory, one can still write
configurations that resemble the gauge fields of magnetic monopoles,
\be
W_i^a = \frac{(1-k(r))}{g\, r}\epsilon^{aij} {\hat x}^j
\ee
where $k(r)$ is a suitable profile function. Can the backreaction due to such
monopole configurations behave like our scalar field $\Phi$ and provide the 
necessary sources for electric strings?

I thank Jude Pereira and Shekhar Sharma for discussions and Gia Dvali for
comments.
This work was supported by the U.S. Department of Energy, Office of High Energy 
Physics, under Award DE-SC0019470 at ASU.

\bibstyle{aps}
\bibliography{paper}

\begin{thebibliography}{14}
\expandafter\ifx\csname natexlab\endcsname\relax\def\natexlab#1{#1}\fi
\expandafter\ifx\csname bibnamefont\endcsname\relax
  \def\bibnamefont#1{#1}\fi
\expandafter\ifx\csname bibfnamefont\endcsname\relax
  \def\bibfnamefont#1{#1}\fi
\expandafter\ifx\csname citenamefont\endcsname\relax
  \def\citenamefont#1{#1}\fi
\expandafter\ifx\csname url\endcsname\relax
  \def\url#1{\texttt{#1}}\fi
\expandafter\ifx\csname urlprefix\endcsname\relax\def\urlprefix{URL }\fi
\providecommand{\bibinfo}[2]{#2}
\providecommand{\eprint}[2][]{\url{#2}}

\bibitem[{\citenamefont{Nielsen and Olesen}(1973)}]{Nielsen:1973cs}
\bibinfo{author}{\bibfnamefont{H.~B.} \bibnamefont{Nielsen}} \bibnamefont{and}
  \bibinfo{author}{\bibfnamefont{P.}~\bibnamefont{Olesen}},
  \bibinfo{journal}{Nucl. Phys. B} \textbf{\bibinfo{volume}{61}},
  \bibinfo{pages}{45} (\bibinfo{year}{1973}).

\bibitem[{\citenamefont{Vilenkin and Shellard}(2000)}]{Vilenkin:2000jqa}
\bibinfo{author}{\bibfnamefont{A.}~\bibnamefont{Vilenkin}} \bibnamefont{and}
  \bibinfo{author}{\bibfnamefont{E.~P.~S.} \bibnamefont{Shellard}},
  \emph{\bibinfo{title}{{Cosmic Strings and Other Topological Defects}}}
  (\bibinfo{publisher}{Cambridge University Press}, \bibinfo{year}{2000}), ISBN
  \bibinfo{isbn}{978-0-521-65476-0}.

\bibitem[{\citenamefont{Schwinger}(1951)}]{Schwinger:1951nm}
\bibinfo{author}{\bibfnamefont{J.~S.} \bibnamefont{Schwinger}},
  \bibinfo{journal}{Phys. Rev.} \textbf{\bibinfo{volume}{82}},
  \bibinfo{pages}{664} (\bibinfo{year}{1951}).

\bibitem[{\citenamefont{Cardona and Vachaspati}(2021)}]{Cardona:2021ovn}
\bibinfo{author}{\bibfnamefont{C.}~\bibnamefont{Cardona}} \bibnamefont{and}
  \bibinfo{author}{\bibfnamefont{T.}~\bibnamefont{Vachaspati}},
  \bibinfo{journal}{Phys. Rev. D} \textbf{\bibinfo{volume}{104}},
  \bibinfo{pages}{045009} (\bibinfo{year}{2021}), \eprint{2105.08782}.

\bibitem[{\citenamefont{Jackiw et~al.}(1979)\citenamefont{Jackiw, Jacobs, and
  Rebbi}}]{Jackiw:1978zi}
\bibinfo{author}{\bibfnamefont{R.}~\bibnamefont{Jackiw}},
  \bibinfo{author}{\bibfnamefont{L.}~\bibnamefont{Jacobs}}, \bibnamefont{and}
  \bibinfo{author}{\bibfnamefont{C.}~\bibnamefont{Rebbi}},
  \bibinfo{journal}{Phys. Rev. D} \textbf{\bibinfo{volume}{20}},
  \bibinfo{pages}{474} (\bibinfo{year}{1979}).

\bibitem[{\citenamefont{Huang and Tipton}(1981)}]{Huang:1980bz}
\bibinfo{author}{\bibfnamefont{K.}~\bibnamefont{Huang}} \bibnamefont{and}
  \bibinfo{author}{\bibfnamefont{R.}~\bibnamefont{Tipton}},
  \bibinfo{journal}{Phys. Rev. D} \textbf{\bibinfo{volume}{23}},
  \bibinfo{pages}{3050} (\bibinfo{year}{1981}).

\bibitem[{\citenamefont{Brown and Weisberger}(1979)}]{Brown:1979bv}
\bibinfo{author}{\bibfnamefont{L.~S.} \bibnamefont{Brown}} \bibnamefont{and}
  \bibinfo{author}{\bibfnamefont{W.~I.} \bibnamefont{Weisberger}},
  \bibinfo{journal}{Nucl. Phys. B} \textbf{\bibinfo{volume}{157}},
  \bibinfo{pages}{285} (\bibinfo{year}{1979}), \bibinfo{note}{[Erratum:
  Nucl.Phys.B 172, 544 (1980)]}.

\bibitem[{\citenamefont{Vachaspati}(2022)}]{Vachaspati:2022ktr}
\bibinfo{author}{\bibfnamefont{T.}~\bibnamefont{Vachaspati}},
  \bibinfo{journal}{Phys. Rev. D} \textbf{\bibinfo{volume}{105}},
  \bibinfo{pages}{105011} (\bibinfo{year}{2022}), \eprint{2204.01902}.

\bibitem[{\citenamefont{Vachaspati and Barriola}(1992)}]{Vachaspati:1992pi}
\bibinfo{author}{\bibfnamefont{T.}~\bibnamefont{Vachaspati}} \bibnamefont{and}
  \bibinfo{author}{\bibfnamefont{M.}~\bibnamefont{Barriola}},
  \bibinfo{journal}{Phys. Rev. Lett.} \textbf{\bibinfo{volume}{69}},
  \bibinfo{pages}{1867} (\bibinfo{year}{1992}).

\bibitem[{\citenamefont{Barriola et~al.}(1994)\citenamefont{Barriola,
  Vachaspati, and Bucher}}]{Barriola:1993fy}
\bibinfo{author}{\bibfnamefont{M.}~\bibnamefont{Barriola}},
  \bibinfo{author}{\bibfnamefont{T.}~\bibnamefont{Vachaspati}},
  \bibnamefont{and} \bibinfo{author}{\bibfnamefont{M.}~\bibnamefont{Bucher}},
  \bibinfo{journal}{Phys. Rev. D} \textbf{\bibinfo{volume}{50}},
  \bibinfo{pages}{2819} (\bibinfo{year}{1994}), \eprint{hep-th/9306120}.

\bibitem[{\citenamefont{Preskill}(1992)}]{Preskill:1992bf}
\bibinfo{author}{\bibfnamefont{J.}~\bibnamefont{Preskill}},
  \bibinfo{journal}{Phys. Rev. D} \textbf{\bibinfo{volume}{46}},
  \bibinfo{pages}{4218} (\bibinfo{year}{1992}), \eprint{hep-ph/9206216}.

\bibitem[{\citenamefont{Workman and Others}(2022)}]{Workman:2022ynf}
\bibinfo{author}{\bibfnamefont{R.~L.} \bibnamefont{Workman}} \bibnamefont{and}
  \bibinfo{author}{\bibnamefont{Others}} (\bibinfo{collaboration}{Particle Data
  Group}), \bibinfo{journal}{PTEP} \textbf{\bibinfo{volume}{2022}},
  \bibinfo{pages}{083C01} (\bibinfo{year}{2022}).

\bibitem[{\citenamefont{Pereira and Vachaspati}(2022)}]{Pereira:2022lbl}
\bibinfo{author}{\bibfnamefont{J.}~\bibnamefont{Pereira}} \bibnamefont{and}
  \bibinfo{author}{\bibfnamefont{T.}~\bibnamefont{Vachaspati}},
  \bibinfo{journal}{Phys. Rev. D} \textbf{\bibinfo{volume}{106}},
  \bibinfo{pages}{096019} (\bibinfo{year}{2022}), \eprint{2207.05102}.

\bibitem[{\citenamefont{Goddard et~al.}(1977)\citenamefont{Goddard, Nuyts, and
  Olive}}]{Goddard:1976qe}
\bibinfo{author}{\bibfnamefont{P.}~\bibnamefont{Goddard}},
  \bibinfo{author}{\bibfnamefont{J.}~\bibnamefont{Nuyts}}, \bibnamefont{and}
  \bibinfo{author}{\bibfnamefont{D.~I.} \bibnamefont{Olive}},
  \bibinfo{journal}{Nucl. Phys. B} \textbf{\bibinfo{volume}{125}},
  \bibinfo{pages}{1} (\bibinfo{year}{1977}).

\end{thebibliography}


\newpage

\appendix

\section{Supplemental Material}

\subsection{Check of solution}
\label{checksoln}

The solution needs to satisfy
\be
(\partial_\nu + g \epsilon^{abc} W_\nu^b)W^{\mu\nu c} = 
i \frac{g}{2} \left ( \Phi^\dag \sigma^a D^\mu\Phi - h.c. \right )
\label{appWeom}
\ee
\be
D^\mu D_\mu \Phi + m^2 \Phi + 2 \lambda | \Phi |^2 \Phi =0 .
\label{appPhieom}
\ee

The solution is
\be
\Phi = \frac{\epsilon}{g} \sqrt{\frac{\Omega}{\omega}} \, f(r)
\begin{pmatrix} z_1 e^{+i\omega t} \\ z_2 e^{-i\omega t} \end{pmatrix}
\label{appstringPhi}
\ee 
and the only non-vanishing components of the gauge field are
\be
W^1_z = -\frac{\epsilon}{g}  \cos( \Omega t ) f(r)  , \ \ 
W^2_z = -\frac{\epsilon}{g}  \sin( \Omega t ) f(r).
\label{appstringW}
\ee
Then
\ba
W_{\mu\nu}^1 &=& -\frac{\epsilon}{g}  \bigl [ - \Omega \sin (\Omega t) f(r) 
(\partial_\mu t \, \partial_\nu z - \partial_\nu t \, \partial_\mu z) \nn \\
&& \hskip 0 cm
+ \cos (\Omega t) f'(r) (\partial_\mu r \, \partial_\nu z - \partial_\nu r \, \partial_\mu z) \bigr ]
\label{appWmunu1}
\ea
\ba
W_{\mu\nu}^2 &=& -\frac{\epsilon}{g}  \bigl [ + \Omega \cos (\Omega t) f(r) 
(\partial_\mu t \, \partial_\nu z - \partial_\nu t \, \partial_\mu z) \nn \\
&& \hskip 0 cm
+ \sin (\Omega t) f'(r) (\partial_\mu r \, \partial_\nu z - \partial_\nu r \, \partial_\mu z) \bigr ]
\label{appWmunu2}
\ea

and $W_{\mu\nu}^3=0$. 
This then gives,
\ba
D_\nu W^{\mu\nu 1} &=& 
-\frac{\epsilon}{g} \cos (\Omega t) \left [ f'' + \frac{f'}{r} +\Omega^2 f \right ] \partial^\mu z
\label{DW1} \\
D_\nu W^{\mu\nu 2} &=& 
-\frac{\epsilon}{g} \sin (\Omega t) \left [ f'' + \frac{f'}{r} +\Omega^2 f \right ] \partial^\mu z 
\label{DW2} \\
D_\nu W^{\mu\nu 3} &=& 
-\frac{\epsilon^2}{g} \Omega f^2 \partial^\mu t \label{DW3}
\ea
where $D_\nu = \partial_\nu + g \epsilon^{abc} W_\nu^b$.

Next we calculate the current
\be
j_\mu^a = i \frac{g}{2} \left ( \Phi^\dag \sigma^a D^\mu\Phi - h.c. \right )
\ee
This calculation is simplified by the identity
\be
\sigma^a \sigma^b = \delta^{ab} +i\epsilon^{abc} \sigma^c
\ee
and by the observation
\be
\partial_\mu\Phi = \left ( \frac{f'}{f} \partial_\mu r + i\omega \sigma_3 \partial_\mu t \right ) \Phi .
\label{partialPhi}
\ee
For then
\be
\Phi^\dag \sigma^1 \partial_\mu \Phi = 
\frac{f'}{f} \partial_\mu r\, (\Phi^\dag \sigma^1 \Phi )
+ \omega \partial_\mu t \, (i\Phi^\dag \sigma^1\sigma^3 \Phi ).
\ee
Both $\Phi^\dag \sigma^1 \Phi$ and $i\Phi^\dag \sigma^1\sigma^3 \Phi =  \Phi^\dag \sigma^2 \Phi $
are real. Therefore
\be
\Phi^\dag \sigma^1 \partial_\mu \Phi - h.c. = 0.
\ee
Similarly
\be
\Phi^\dag \sigma^2 \partial_\mu \Phi - h.c. = 0
\ee
and
\be
\Phi^\dag \sigma^3 \partial_\mu \Phi - h.c. = 2i\omega |\Phi|^2 \partial_\mu t
\ee
Similar considerations apply to $\Phi^\dag \sigma^a W_\mu^b \sigma^b \Phi$ and we
get
\ba
&&
\hskip -8 cm
-i \frac{g}{2} \Phi^\dag \sigma^a W_\mu^b \sigma^b \Phi - h.c. =
\nn \\ 
\left ( -i \frac{g}{2} W_\mu^a |\Phi|^2 + 
\frac{g}{2} \epsilon^{abc} W_\mu^b \Phi^\dag \sigma^c \Phi \right ) - h.c.
= -i 2g W_\mu^a |\Phi|^2 \nn
\ea
For now we consider $\omega$ to be positive or negative. Then,
\be
|\Phi |^2 = \frac{\epsilon^2}{g^2} \frac{\Omega}{|\omega |} f^2 .
\ee
The currents from the $\Phi$ field are
\ba
j_\mu^1 &=& 
\frac{g^2}{2}W_\mu^1 |\Phi |^2 \nn \\
&&
= -\frac{\epsilon}{g}\cos( \Omega t ) \Omega^2 \frac{\epsilon^2}{2|\omega |\Omega}  f^3
\partial_\mu z
\label{jmu1} \\
j_\mu^2 &=& 
\frac{g^2}{2}W_\mu^1 |\Phi |^2 \nn \\
&&
= -\frac{\epsilon}{g}\sin( \Omega t )  \Omega^2 \frac{\epsilon^2}{2|\omega |\Omega}  f^3
\partial_\mu z
\label{jmu2} \\
j_\mu^3 &=& -\frac{\epsilon^2}{g} \frac{\omega}{|\omega |} \Omega f^2 \partial_\mu t 
\label{jmu3}
\ea
Matching \eqref{jmu3} to \eqref{DW3} we see that it is necessary to restrict $\omega$
to be positive and hence we only consider $\omega > 0$ below. Further, matching
\eqref{DW1}-\eqref{DW2} with \eqref{jmu1}-\eqref{jmu2}, we see that the
gauge equations are satisfied if $f(r)$ is a solution of 
\be
 f'' + \frac{f'}{r} + \Omega^2 \left ( 1 - \frac{\epsilon^2 }{2\omega  \Omega} f^2 \right ) f = 0.
 \label{appfeq}
\ee

Next we come to the $\Phi$ equation of motion in \eqref{appPhieom}.
\ba
D^\mu D_\mu \Phi &=& 
\left (\partial^\mu -i\frac{g}{2} W^{\mu a}\sigma^a \right ) 
\left (\partial_\mu -i\frac{g}{2} W_\mu^b\sigma^b \right)\Phi \nn \\
&=& \partial^\mu \partial_\mu \Phi - \frac{g^2}{4} W^{\mu a} W_\mu^b 
(\delta^{ab} +i \epsilon^{abc} \sigma^c )\Phi \nn \\
&=& \partial^\mu \partial_\mu \Phi - \frac{g^2}{4} W^{\mu a} W_\mu^a \Phi \nn
\ea
where we have used $\partial_\mu W^{\mu a}=0 = W^{\mu a} \partial_\mu \Phi$,
and $\epsilon^{abc}W^{\mu a} W_\mu^b =0$. To simplify further, we use
\eqref{partialPhi} to get,
\ba
 \partial^\mu \partial_\mu \Phi &=& \partial^\mu \left ( \frac{f'}{f} \partial_\mu r \Phi \right )
  +  \left ( \frac{f'}{f} \partial_\mu r + i\omega \sigma_3 \partial_\mu t \right )^2 \Phi \nn \\
 &=& - \left ( \frac{f''}{f} + \frac{f'}{rf}  + \omega^2 \right ) \Phi
\ea
Therefore the equation of motion for $\Phi$ becomes
\ba
0 &=& D^\mu D_\mu \Phi + m^2 \Phi + 2 \lambda | \Phi |^2 \Phi \nn \\
&=& 
- \left ( f'' + \frac{f'}{r}  + \omega^2 f - \frac{\epsilon^2}{4}  f^3 
-m^2 f - 2\lambda \frac{\epsilon^2}{g^2}\frac{\Omega}{\omega} f^3 \right ) \frac{\Phi}{f} \nn
\ea
or
\be
f'' + \frac{f'}{r}  + (\omega^2-m^2) f - 
\left (\frac{\epsilon^2}{4}  + 2\lambda \frac{\epsilon^2}{g^2}\frac{\Omega}{\omega} \right ) f^3 =0
\label{feqfromPhi}
\ee
From \eqref{Omsol} and \eqref{omsol} one checks that 
\be
\omega^2-m^2 = \Omega^2
\ee
and
\ba
\frac{\epsilon^2}{4}  + 2\lambda \frac{\epsilon^2}{g^2}\frac{\Omega}{\omega}  &=&
\frac{\epsilon^2}{4} \left ( 1 + \frac{8\lambda}{g^2} \frac{\Omega}{\omega} \right ) \nn \\
&=&
\frac{\epsilon^2}{4} \left ( 1 + \frac{4\lambda}{g^2} \frac{1}{1-4\lambda/g^2} \right ) \nn \\
&=& \frac{\epsilon^2}{4(1-4\lambda/g^2)} \nn \\
&=& \frac{\epsilon^2 \Omega^2}{2 \omega \Omega}
\ea
Using these relations, \eqref{feqfromPhi} is identical to \eqref{appfeq} and so the $\Phi$
equation is satisfied too.

\end{document}